\documentclass[twocolumn,aps,prd,10pt,superscriptaddress,longbibliography,floatfix,citeautoscript]{revtex4-2}

\usepackage{amsmath}
\usepackage{amssymb}
\usepackage{bm}
\usepackage{dcolumn}
\usepackage{glossaries}
\usepackage{graphicx}
\usepackage{multirow}
\usepackage{fancyvrb}
\usepackage[linesnumbered,ruled]{algorithm2e}
\usepackage{siunitx}

\usepackage[usenames,dvipsnames,svgnames]{xcolor}
\usepackage{hyperref}
\hypersetup{
    pdfnewwindow=true,      
    colorlinks=true,        
    linkcolor=Blue,         
    citecolor=Blue,         
    filecolor=Blue,         
    urlcolor=Blue           
}

\usepackage{listings}	
\newacronym{dft}{DFT}{density functional theory}
\newacronym{emd}{EMD}{equilibrium molecular dynamics}
\newacronym{hnemd}{HNEMD}{homogeneous non-equilibrium molecular dynamics}
\newacronym{md}{MD}{molecular dynamics}
\newacronym{mlp}{MLP}{machine-learned potential}
\newacronym{nep}{NEP}{neuroevolution potential}
\newacronym{rmse}{RMSE}{root mean square error}
\newacronym{qhp}{qHP}{quasi-hexagonal phase}
\newacronym{qtp}{qTP}{quasi-tetragonal phase}
\newacronym{2d}{2D}{two-dimensional}
\newacronym{rdf}{RDF}{radial distribution function} 
\newacronym{adf}{ADF}{angular distribution function}
\newacronym{mfp}{MFP}{mean free path}

\DeclareSIUnit\angstrom{\text{Å}}
\DeclareSIUnit{\atom}{atom}
\DeclareSIUnit{\step}{step}
\DeclareSIUnit{\atomstepsecond}{\atom\step\per\second}

\begin{document}


\title{Anisotropic and isotropic elasticity and thermal transport in monolayer C$_{24}$ networks from machine-learning molecular dynamics}

\author{Qing Li}
\affiliation{College of Physical Science and Technology, Bohai University, Jinzhou, 121013, China}
\author{Haikuan Dong}
\email{donghaikuan@163.com}
\affiliation{College of Physical Science and Technology, Bohai University, Jinzhou, 121013, China}
\author{Penghua Ying}
\email{hityingph@tauex.tau.ac.il}
\affiliation{Department of Physical Chemistry, School of Chemistry, Tel Aviv University, Tel Aviv, 6997801, Israel}
\author{Zheyong Fan}
\email{brucenju@gmail.com}
\affiliation{College of Physical Science and Technology, Bohai University, Jinzhou, 121013, China}
\date{\today}

\begin{abstract}
Two-dimensional fullerene networks have recently attracted increasing interest due to their diverse bonding topologies and mechanically robust architectures. 
In this work, we develop an accurate machine-learned potential NEP-C$_{24}$ for both the quasi-hexagonal phase (qHP) and the quasi-tetragonal phase (qTP) C$_{24}$ monolayers, based on the neuroevolution potential (NEP) framework. 
Using this NEP-C$_{24}$ model, we systematically investigate the elastic and thermal transport properties. 
Compared with C$_{60}$ monolayers, both C$_{24}$ phases exhibit markedly enhanced stiffness, arising from the combination of reduced molecular size and increased density of covalent bonds. 
The qTP C$_{24}$ monolayer shows nearly isotropic elastic properties and thermal conductivities along its two principal axes owing to its four-fold symmetry, whereas the chain-like, misaligned bonding topology of the qHP C$_{24}$ monolayer leads to pronounced in-plane anisotropy. 
Homogeneous nonequilibrium molecular dynamics and spectral decomposition analyses reveal that low-frequency ($<5$ THz) acoustic phonons dominate heat transport, with directional variations in phonon group velocity and mean free path governing the anisotropic response in qHP C$_{24}$. 
Real-space heat flow visualizations further show that, in these fullerene networks, phonon transport is dominated by strong inter-fullerene covalent bonds rather than weak van der Waals interactions.
These findings establish a direct link between intermolecular bonding topology and phonon-mediated heat transport, providing guidance for the rational design of fullerene-based two-dimensional materials with tunable mechanical and thermal properties.
\end{abstract}

\maketitle

\section{Introduction}
Carbon is fundamental to life on Earth and displays exceptional versatility owing to its diverse hybridization states. 
Its ability to form sp, sp$^2$, and sp$^3$ bonds enables a variety of polymorphs. 
For example, the sp$^3$-hybridized form constitutes the cubic lattice of diamond, one of the hardest known materials, while sp$^2$-hybridized carbon forms layered graphite with weak interlayer van der Waals interactions that underlie its lubricating properties~\cite{hod2018structural, ying2025scaling}. 
Materials with hybrid sp$^3$- and sp$^2$-bonds also exist, as evidenced by the recently synthesized fullerene networks based on C$_{60}$~\cite{hou2022nature}.
The physical and chemical properties of the C$_{60}$ networks have been actively studied \cite{dong2023ijhmt, ying2023extreme,peng2023nanolett,HU2025ijhmt,YANG2024ijhmt,ying2024superlubric}.

Apart from C$_{60}$, other fullerene molecules can also form 2D networks. 
In particular, C$_{24}$ is known as the smallest stable fullerene molecule \cite{kroto1985nature,cox1986jacs,kroto1987nature}.
Recently, the electronic structures of C$_{24}$ networks have been studied using first principles calculations, showing that these networks are semiconductors~\cite{wu2025jacs}.
However, a fundamental understanding of their mechanical robustness and phonon-mediated heat transport, particularly the interplay between bonding topology and anisotropic transport remains largely unexplored.

An in-depth investigation of the mechanical and thermal transport properties of these 2D networks involves large spatial and time scales, for which \gls{md} is a feasible approach.
A crucial ingredient of \gls{md} simulations is the underlying interatomic potential. 
For carbon systems, several empirical potentials, such as the Tersoff potential~\cite{tersoff1989prb, lindsay2010prb}, have been widely used.
However, they are primarily parameterized for diamond or graphene, which have difficulty in describing the fullerene networks with mixed sp$^2$ and sp$^3$ bonds.
Recently, \glspl{mlp} have emerged as a powerful and flexible approach for accurately describing the potential energy surface of diverse materials~\cite{behler2007prl}. 
Among various \gls{mlp} frameworks, the \gls{nep} method~\cite{fan2021neuroevolution,fan2022jpcm,fan2022jcp} stands out for its high computational efficiency.
\gls{nep} has been widely used to investigate mechanical \cite{ying2023extreme, BU2025mtp,zhao2024prb} and thermal transport properties~\cite{dong2023ijhmt, dong2024jap, LI2025MTP,WU2024MTP, ZHOU2025MTP, xu2025npjcompmat} across diverse material systems. 
Particularly, some of the present authors applied the \gls{nep} method to investigate the mechanical and thermal transport properties of \gls{qhp} C$_{60}$~\cite{ying2023extreme, dong2023ijhmt}, demonstrating its reliability in describing the thermodynamic behavior of fullerene network systems.

In this work, we develop an accurate \gls{nep} model capable of simultaneously describing the \gls{qhp} and \gls{qtp} C$_{24}$ monolayers, and systematically study their elastic and thermal transport properties using static and dynamic calculations.
For elasticity, we calculate the elastic constants, Young’s modulus, shear modulus, and Poisson’s ratio to assess the mechanical stability and anisotropy. 
For thermal transport, we focus on the lattice thermal conductivity as these materials are semiconducting.
Our results establish clear structure–property relationships linking intermolecular bonding motifs to directional elasticity and phonon thermal transport.
These insights provide theoretical guidance for engineering fullerene-based \gls{2d} materials with customized heat conduction characteristics.

\begin{figure}[htbp]  
    \centering
    \includegraphics[width=0.5\textwidth]{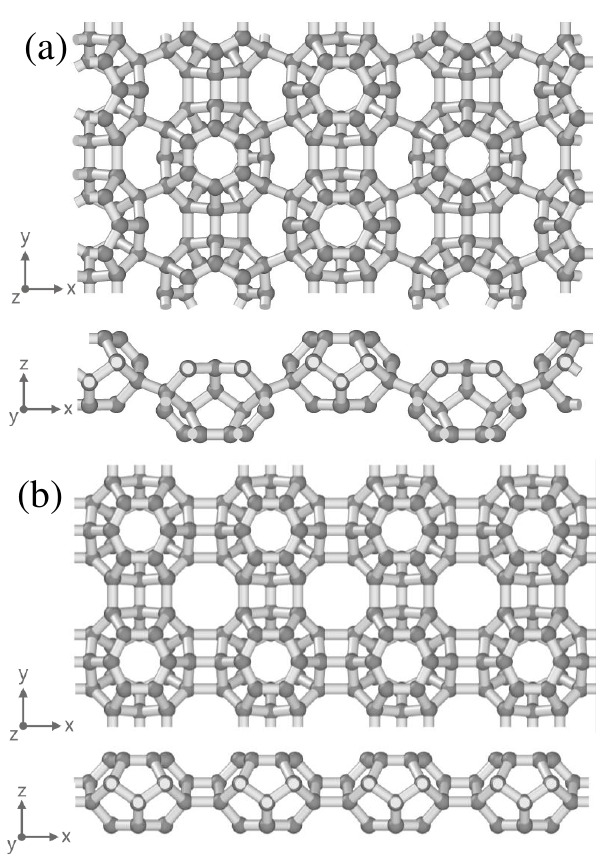} 
    \caption{Top and side views of the optimized crystal structures of (a) \gls{qhp} and (b) \gls{qtp} C$_{24}$ monolayers. The structures are visualized using OVITO~\cite{Stukowski2009msme}.}
    \label{figstrucure}  
\end{figure} 
 
\section{Models and methods}
\subsection{The crystal structures of quasi-hexagonal and quasi-tetragonal phases}

The initial crystal structures of the \gls{qhp} and \gls{qtp} C$_{24}$ monolayers are shown in \autoref{figstrucure}. 
In \gls{qhp} monolayer (\autoref{figstrucure}(a)), each C$_{24}$ molecule is covalently bonded to six neighboring C$_{24}$ molecules, forming an extended two-dimensional network. 
The \gls{qhp} C$_{24}$ monolayer can be regarded as a network of misaligned one-dimensional chains oriented along the $y$-direction and interconnected by three noncoplanar bonds, where the central bond is slightly shorter than the two lateral ones. 
Adjacent chains are further linked by diagonal single bonds along the $x$-direction. 
Owing to the asymmetric interchain bonding, the monolayer adopts a buckled structure.

The overall structure of the \gls{qtp} C$_{24}$ monolayer (\autoref{figstrucure}(b)) resembles that of its C$_{60}$ counterpart, forming a nearly square lattice stabilized by intermolecular covalent bonds between adjacent C$_{24}$ molecules. 
Each C$_{24}$ molecule is covalently connected to four neighboring C$_{24}$ molecules, constructing an extended two-dimensional framework. 
Moreover, unlike the [2 + 2] cycloaddition bonds observed in the \gls{qtp} C$_{60}$ monolayer, neighboring C$_{24}$ clusters in \gls{qtp} C$_{24}$ are connected through three noncoplanar bonds, with the central bond being marginally shorter than the two lateral ones. 

\subsection{The NEP approach for machine-learned potential}
\gls{nep} is a type of \gls{mlp} that combines an evolution strategy~\cite{schaul2011gecco} with a feed-forward neural network~\cite{fan2021neuroevolution, fan2022jpcm, fan2022jcp}.
Adopting the conventional Behler-Parrinello~\cite{behler2007prl} high-dimensional neural network methodology, the site energy $U_i$ of an atom $i$ is defined as a function of $N_{\text{des}}$ descriptor components, with its form given below:

\begin{equation}
\label{eq:energy_potential} 
U_i = \sum_{\mu=1}^{N_{\text{neu}}} w_{\mu}^{(1)} \tanh\left( \sum_{\nu=1}^{N_{\text{des}}} w_{\mu\nu}^{(0)} q_{\nu}^i - b_{\mu}^{(0)} \right) - b^{(1)},
\end{equation}
where $N_{\text{neu}}$ denotes the number of neurons, $\boldsymbol{w}^{(0)}$ and $\boldsymbol{w}^{(1)}$ represent the trainable weights, $\boldsymbol{b}^{(0)}$ and $\boldsymbol{b}^{(1)}$ stand for the bias parameters, and $\tanh(x)$ serves as the activation function.

In the \gls{nep} method, the descriptor is composed of a set of radial and angular components. The radial descriptor components $q_n^i$ are formulated as:
\begin{equation}
\label{equation:qin}
q^i_{n}
= \sum_{j\neq i} g_{n}(r_{ij})
\quad\text{with}\quad
0\leq n\leq n_\mathrm{max}^\mathrm{R},
\end{equation}
where the summation includes all neighboring atoms of atom $i$ located within the specified cutoff distance.

For the angular descriptor components, we take into account both three-body and four-body ones. 
For 3-body ones ($0\leq n\leq n_\mathrm{max}^\mathrm{A}$, $1\leq l \leq l_\mathrm{max}^\mathrm{3b}$)
\begin{equation}
q^i_{nl} 
= \sum_{m=-l}^l (-1)^m A^i_{nlm} A^i_{nl(-m)},
\label{equation:qinl_spherical}
\end{equation}
and 4-body ones can be found in Ref.~\citenum{fan2022jcp}.
Here,
\begin{equation}
A^i_{nlm} 
= \sum_{j\neq i} g_n(r_{ij}) Y_{lm}(\theta_{ij},\phi_{ij}),
\label{equation:Ainlm}
\end{equation}
and \( Y_{lm}(\theta_{ij}, \phi_{ij}) \) denotes the spherical harmonics, which depend on the polar angle \( \theta_{ij} \) and the azimuthal angle \( \phi_{ij} \) corresponding to the position difference \( \boldsymbol{r}_{ij} = \boldsymbol{r}_j - \boldsymbol{r}_i \) between atom \( i \) and atom \( j \). 

The radial functions \( g_n(r_{ij}) \) in \autoref{equation:qin} are defined as a linear combination of \( N_{\text{bas}}^{\text{R}} + 1 \) basis functions:
\begin{equation}
g_n(r_{ij}) = \sum_{k=0}^{N_{\text{bas}}^{\text{R}}} c_{nk}^{ij} f_k(r_{ij}), 
\end{equation}
with
\begin{equation}
f_k(r_{ij}) = \frac{1}{2} \left[ T_k \left( 2 \left( r_{ij}/r_{\text{c}}^{\text{R}} - 1 \right)^2 - 1 \right) + 1 \right] f_{\text{c}}(r_{ij}).
\end{equation}
Here, \( T_k(x) \) denotes the Chebyshev polynomial of the first kind of order $k$, and \( f_{\text{c}}(r_{ij}) \) represents the cutoff function defined as:
\begin{equation}
f_{\text{c}}(r_{ij}) =
\begin{cases}
\frac{1}{2} \left[ 1 + \cos \left( \pi \frac{r_{ij}}{r_{\text{c}}^{\text{R}}} \right) \right], & r_{ij} \leq r_{\text{c}}^{\text{R}}; \\
0, & r_{ij} > r_{\text{c}}^{\text{R}}.
\end{cases}
\end{equation}
Here, \( r_{\text{c}}^{\text{R}} \) represents the cutoff distance for the radial descriptor components. The trainable expansion coefficients \( c_{nk}^{ij} \) are dependent on \( n \), \( k \), as well as the types of atoms \( i \) and \( j \). The functions \( g_n(r_{ij}) \) in \autoref{equation:qinl_spherical}  are defined similarly, with the exception of a different basis size \( N_{\text{bas}}^{\text{A}} \) and a different cutoff distance \( r_{\text{c}}^{\text{A}} \).

The descriptor components are assembled into a vector $\{q^i_\nu\}_{\nu=1}^{N_{\text{des}}}$ of dimension $N_{\text{des}}$, which serves as the input layer of a feedforward neural network comprising a single hidden layer with $N_{\text{neu}}$ neurons. The network output corresponds to the potential energy of atom $i$.

\subsection{Training the NEP model}
A general-purpose \gls{nep} model for carbon systems has been recently developed~\cite{fan2024jphys}; however, its training dataset does not explicitly include C$_{24}$ structures~\cite{deringer2017prb}. Although both the \gls{qhp} and \gls{qtp} monolayers remain stable in \gls{md} simulations using this model, its accuracy for these structures is insufficient, as demonstrated later. To achieve higher fidelity in \gls{md} simulations, we constructed a new \gls{nep} model trained on \gls{qhp} and \gls{qtp} structures with \gls{dft} reference data. For clarity, the previously published and newly developed models are referred to as \gls{nep}-Carbon and \gls{nep}-C$_{24}$, respectively.
\subsubsection {Generation of training and testing structures}
The training and testing datasets include both \gls{qhp} and \gls{qtp} structures. For each phase, \gls{md} simulations were performed using the \gls{nep}-Carbon~\cite{fan2024jphys} model within the NVT ensemble. A rectangular simulation cell containing 96 carbon atoms was adopted for both the \gls{qhp} and \gls{qtp} monolayers, with the Langevin thermostat applied to maintain temperature control.

For the \gls{qhp} monolayer, we considered seven uniaxial strain conditions: $-3\%$, $-2\%$, $-1\%$, $0\%$, $+1\%$, $+2\%$ and $+3\%$ applied independently in both the $x$ and $y$ directions. Under each strain, the temperature was linearly increased from 100 K to 1000 K over a 1000 ps simulation. We uniformly selected 25 structures from the trajectory for each target strain, yielding a total of 325 \gls{qhp} structures.
The \gls{qtp} monolayer was treated following the same simulation protocol, except that strain was applied only along one lateral direction due to its structural isotropy. This yielded 175 \gls{qtp} structures. In total, 500 structures were obtained, among which 400 (38,400 atoms) were randomly selected for training and 100 (9,600 atoms) for testing.

\subsubsection{DFT calculations}
After we obtained these refrence structures, we applied quantum-mechanical \gls{dft} calculations to obtain their reference energy, force, and virial data. To this end, we used the PBE functional~\cite{Perdew1996prl} combined with the many-body dispersion correction~\cite{Tkatchenko2012prl} as implemented in the VASP package~\cite{Kresse1996prb,kresse1996computational}. An energy cutoff of 650 ev was applied for the projector augmented wave~\cite{paw1,paw2}. A G-centered k-point mesh corresponding to a density of 0.2 $\rm \AA^{-1}$ and an electronic convergence criterion of $10^{-7}$ eV were used in the self-consistent calculations. A Gaussian smearing with a width of 0.1 eV was employed.

\begin{table}[h]
    \centering
    \caption{Hyperparameters for the \gls{nep}-C$_{24}$ model.}
    \label{tab:NEP_C24}
    \renewcommand{\arraystretch}{1.3}
    \begin{tabular*}{\columnwidth}{@{\extracolsep{\fill}}l c l c @{}}
        \hline
        Parameter & Value & Parameter & Value \\
        \hline
        $r_{\rm c}^{\rm R}$ & $7 \mathring{\text{A}}$ & $r_{\rm c}^{\rm A}$ & $4 \mathring{\text{A}}$ \\
        $n_{\rm{max}}^{\rm R}$ & 8 & $n_{\rm{max}}^{\rm A}$ & 8 \\
        $N_{\rm{bas}}^{\rm R}$ & 12 & $N_{\rm{bas}}^{\rm A}$ & 12 \\
        $l_{\rm{max}}^{3\rm{b}}$ & 4 & $l_{\rm{max}}^{4\rm{b}}$ & 2 \\
        $N_{\rm{neu}}$ & 50 &  $\lambda_{\rm e}$ & 1.0 \\
        $\lambda_{\rm f}$ & 1.0 & $\lambda_{\rm v}$ & 0.1 \\
        $N_{\rm{bat}}$ & 10000 & $N_{\rm{pop}}$ & 50 \\
        $N_{\rm{gen}}$ & $3 \times 10^5$ &  &  \\
        \hline
    \end{tabular*}
\end{table}

\subsubsection{Choosing the training hyperparameters}
We employed the GPUMD package~\cite{xu2025mgadvances} to train the \gls{nep}-C$_{24}$ model with the hyperparameters listed in \autoref{tab:NEP_C24}. Compared to the hyperparameter values for the \gls{nep}-Carbon model~\cite{fan2024jphys}, several modifications were introduced. First, the five-body descriptor components defined in Ref~\cite{fan2022jcp} were not included. Second, we have decreased $n_{\rm{max}}^{\rm{R}}$ from 12 to 8 and $N_{\rm{bas}}^{\rm{R}}$ from 16 to 12 to control model complexity without sacrificing accuracy.

\subsection{The homogeneous nonequilibrium molecular dynamics method}

We computed the thermal conductivity using the \gls{hnemd} method. 
This method was initially developed for two-body potentials~\cite{evans1982pla} and has been extended to many-body potentials~\cite{Fan2019prb}, including \glspl{mlp} with atom-centered descriptors~\cite{fan2021neuroevolution}. 
In this method, we apply an external driving force with a net value of zero to the atoms in the system. 
This induces a heat current that exhibits a nonzero ensemble average  $\langle\mathbf{J}\rangle$. In the linear-response regime, the heat current is proportional to the driving force parameter $\mathbf{F}_{\rm e}$:
\begin{equation}
\label{equation:J}
\langle J^{\alpha} \rangle = T V \sum_{\beta} \kappa^{\alpha\beta} F_{\rm e}^{\beta},
\end{equation}
Here, $T$ represents the temperature and $V$ denotes the volume in the system. The proportionality constant $\kappa^{\alpha\beta}$ corresponds to the $\alpha\beta$-th component of the thermal conductivity tensor. In our study, we focused exclusively on the diagonal principal components of this tensor. For a specific direction $\alpha$, the thermal conductivity component is thus defined as $\kappa^\alpha \equiv \kappa^{\alpha\alpha}$ and calculated as follows:
\begin{equation}
\label{equation:kappa}
\kappa^{\alpha} = \frac{\langle J_{\alpha}\rangle}{TVF_{\rm e}^{\alpha}}.
\end{equation}

The heat current can be decomposed in the frequency domain, thus obtaining the spectral thermal conductivity~\cite{Fan2019prb}:
\begin{equation}
    \kappa^{\alpha}(\omega) = \frac{2}{VTF_{\rm e}^{\alpha}}\int_{-\infty}^{\infty} \text{d}t e^{\text{i}\omega t} \sum_i\sum_{j\neq i} \left\langle r^{\alpha}_{ij} \frac{\partial U_j}{\partial \bm{r}_{ji}}(0) \cdot \bm{v}_i(t) \right\rangle.
\end{equation}
Here, $U_j$ is the site energy of atom $j$, $\bm{r}_{ji}=\bm{r}_{i}-\bm{r}_{j}$, $\bm{r}_{i}$ is the position of atom $i$, and $\bm{v}_{i}$ is the velocity of atom $i$.

\begin{figure*}
    \centering
    \includegraphics[width=1\textwidth]{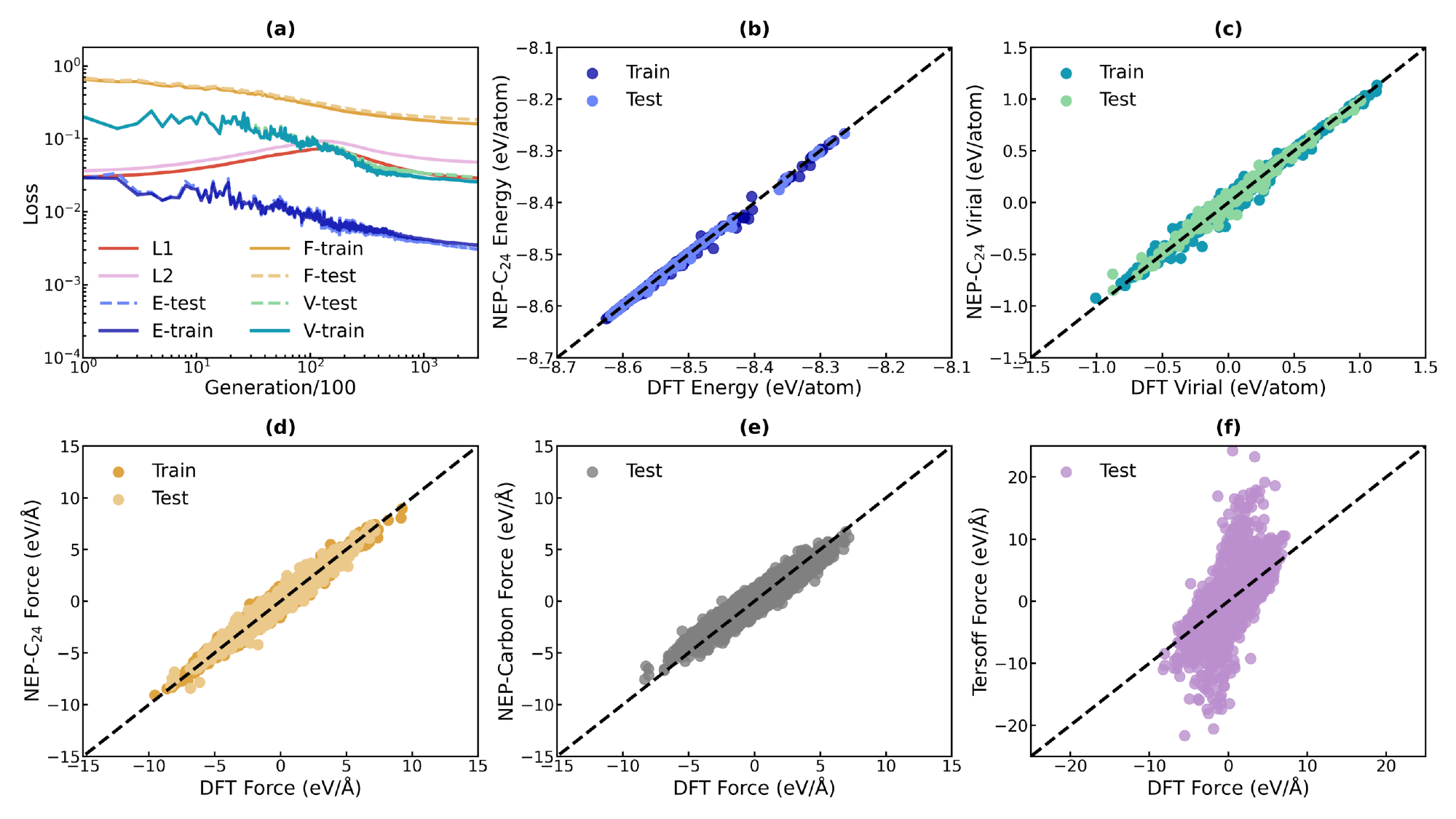}  
    \caption{(a) Evolution of the individual components of the loss function for the training and testing datasets as a function of generation. (b–d) Comparison of (b) energy, (c) virial, and (d) force predicted by \gls{nep}-C$_{24}$ with \gls{dft} reference data for both training and testing datasets. (e,f) Comparison of the forces obtained from (e) \gls{nep}-Carbon and (f) Tersoff potentials with \gls{dft} reference data for the testing dataset.}
    \label{fignep}  
\end{figure*} 

\section{Results and discussion}
\subsection{NEP validation}
Figure~\ref{fignep} shows the evolution of the various components in the loss function during the training process. After 3×$10^{5}$ generations, all loss terms converge, and  the predicted energy, virial, and force by the \gls{nep}-C$_{24}$ model show excellent agreement with \gls{dft} reference data (\autoref{fignep} (b)–(d)). For the training dataset, the overall \gls{rmse} values for energy, virial, and force are 3.5 meV/atom, 25.9 meV/atom, and 161.0 meV/\AA, respectively. For the test dataset, the corresponding \gls{rmse} values are 3.1 meV/atom, 29.9 meV/atom, and 184.2 meV/\AA. For comparison, in \autoref{fignep} (e)-(f), we also evaluate the accuracy of the previous \gls{nep}-Carbon model~\cite{fan2024jphys} and widely used Tersoff potential for the graphene-like systems~\cite{lindsay2010prb} in reproducing the DFT forces of the test dataset. The \gls{nep}-C$_{24}$ model achieves a force \gls{rmse} of 184.2 meV/\AA, substantially lower than 520.8 meV/$\rm \AA$ for \gls{nep}-Carbon and approximately 2000 meV/$\rm \AA$ for Tersoff, demonstrating highest accuracy among all the tested potentials. 

\begin{figure}[htbp]
    \centering
    \includegraphics[width=0.5\textwidth]{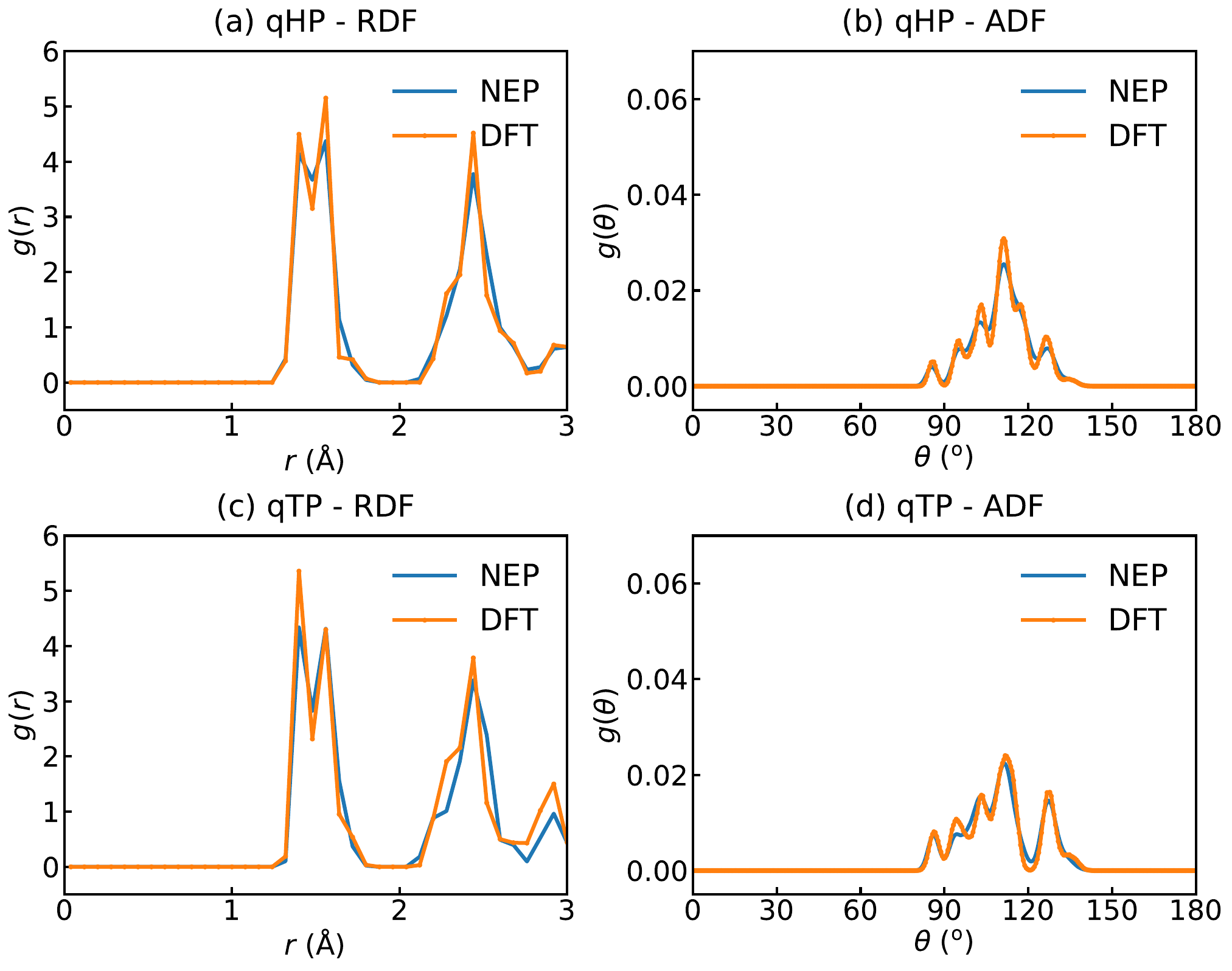} 
    \caption{(a,c) \glspl{rdf} and (b,d) \glspl{adf} of (a,b) \gls{qhp} and (c,d) \gls{qtp} C$_{24}$ monolayer, obtained from the \gls{nep}-$\rm C_{24}$- and \gls{dft}-based \gls{md} simulations at 300 K. \gls{dft}-\gls{md} simulations were performed for 10 ps using cells containing 48 atoms (\gls{qhp}) and 24 atoms (\gls{qtp}), while \gls{nep}-MD simulations were performed for 10 ns using systems containing 31,920 atoms (\gls{qhp}) and 38,400 atoms (\gls{qtp}).}
    \label{figrdf} 
\end{figure} 

\begin{figure}[htbp]
    \centering
    \includegraphics[width=0.5\textwidth]{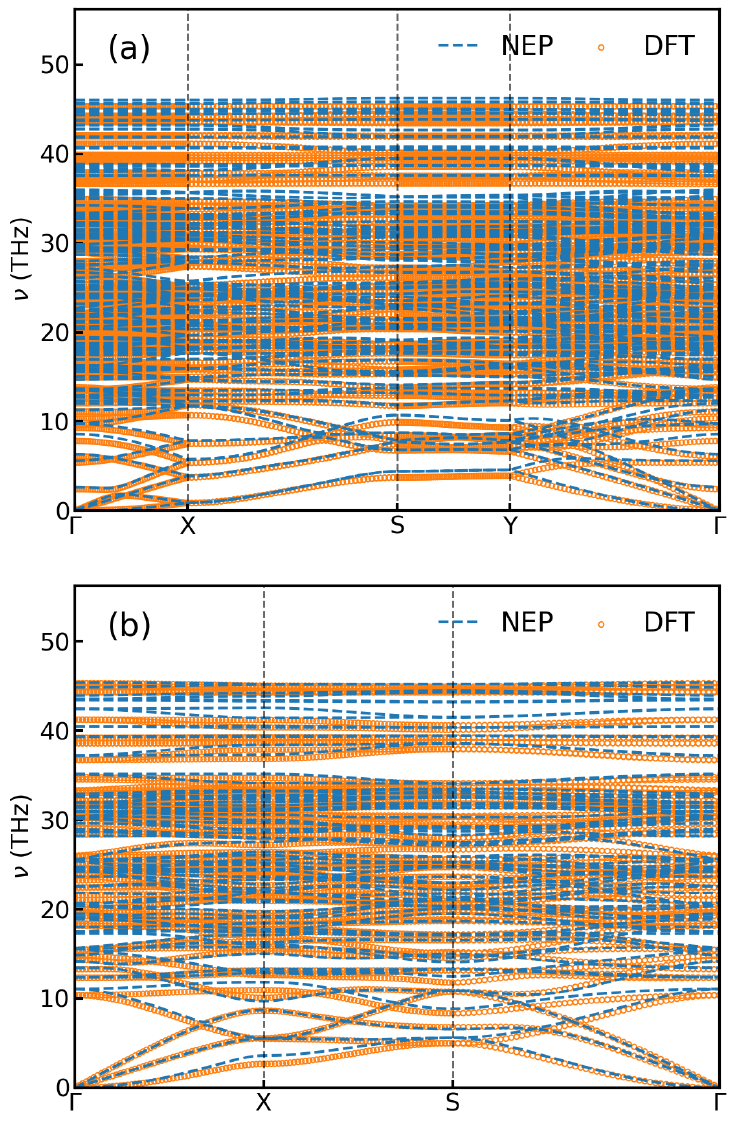} 
    \caption{Phonon dispersions of (a) \gls{qhp} and (b) \gls{qtp} C$_{24}$ monolayers calculated using \gls{nep}-C$_{24}$ and \gls{dft}. The DFT results are consistent with those reported in Ref.~\cite{wu2025jacs}}.
    \label{figphonon} 
\end{figure} 

Beyond \gls{rmse} metrics, we also calculated the \gls{rdf} and \gls{adf} of the \gls{qhp} and \gls{qtp} structures at 300 K as an additional validation. Fig.~\ref{figrdf} presents the comparison between the \gls{nep}-C$_{24}$ and \gls{dft} results. For both phases, the \glspl{rdf} and \glspl{adf} predicted by \gls{nep} are in good agreement with the \gls{dft} reference results, with slight deviations likely arising from the difference in simulation sizes and sampling times. This demonstrates the reliability of \gls{nep}-C$_{24}$ in describing short-range order and local atomic environments in both phases.

In addition, we calculated the phonon dispersion relations of both \gls{qhp} and \gls{qtp} structures, as shown in \autoref{figphonon} (a) and (b), respectively. The \gls{nep}-C$_{24}$ predictions are in excellent agreement with \gls{dft} results, indicating that the developed potential reliably captures the harmonic interatomic interactions in C$_{24}$ systems and thus provides a solid basis for subsequent investigations of phonon transport. Near the $\Gamma$ point, the dispersion curves for both \gls{qhp} and \gls{qtp} structures exhibit two Debye-like linear acoustic branches corresponding to the in-plane longitudinal and transverse modes, as well as a quadratic acoustic branch associated with the out-of-plane flexural mode, a characteristic feature of monolayer systems. No imaginary-frequency phonon modes are observed in either phase, indicating that the structures lie at local minima on the potential energy surface and are stable at zero temperature.

\subsection{Elastic properties}
We next investigate the elastic properties of the two monolayer phases. The elastic constants of the \gls{qhp} and \gls{qtp} C$_{24}$ monolayers were obtained with the aid of the VASPKIT package~\cite{wang2021cpc}. Each component of the elastic tensor was derived from the first-order derivative of the stress-strain relationship, evaluated over five small strain values ranging from $-2\%$ to $2\%$ within the elastic limit. The four independent elastic constants for each phase predicted by  \gls{nep}-C$_{24}$ and \gls{dft} are listed in \autoref{table:Cij}. The \gls{nep} results are in close agreement with the \gls{dft} results, further confirming the accuracy of the constructed \gls{nep}-C$_{24}$ for describing the elastic response of the two monolayers.

\begin{table*}[htb]
\centering
\setlength{\tabcolsep}{2.5mm}
\caption{Elastic constants $C_{ij}$ (N/m) of monolayer carbon allotropes.}
\label{table:Cij}
\begin{tabular}{llllll} %
\hline
Phase & Method& C$_{11}$ & C$_{22}$ & C$_{12}$ & C$_{66}$ \\
\hline
\gls{qhp} C$_{24}$ & \gls{dft} (This work) & 225.1 & 275.7 & 23.6 & 106.0  \\
\gls{qhp} C$_{24}$ & \gls{nep} (This work)  & 236.5 & 295.4 & 24.4 & 110.3  \\
\gls{qtp} C$_{24}$ & \gls{dft} (This work)  & 251.3 & 251.3 & -4.4 & 79.9   \\
\gls{qtp} C$_{24}$ & \gls{nep} (This work)  & 248.8 & 248.8 & -15.8 & 81.6   \\
\gls{qhp} C$_{60}$& \gls{dft}~\cite{peng2023nanolett}  & 150.8 & 186.8 & 22.5   & 60.6  \\
\gls{qtp}2 C$_{60}$& \gls{dft}~\cite{peng2023nanolett} & 149.9 & 148.7 & 22.9   & 53.4  \\
graphene & \gls{dft}~\cite{Liu2021PCCP} &351.4 &351.4 & 61.6
& 144.9 \\
$\alpha$-graphyne & \gls{dft}~\cite{hou2018study} & 95 & 95 & 82 & 6.5   \\
biphenylene & \gls{dft}~\cite{luo2021scirep} &294 & 240 & 91& 83  \\

\hline
\end{tabular}
\end{table*}

\begin{figure*}[htb]
    \centering
    \includegraphics[width=1\textwidth]{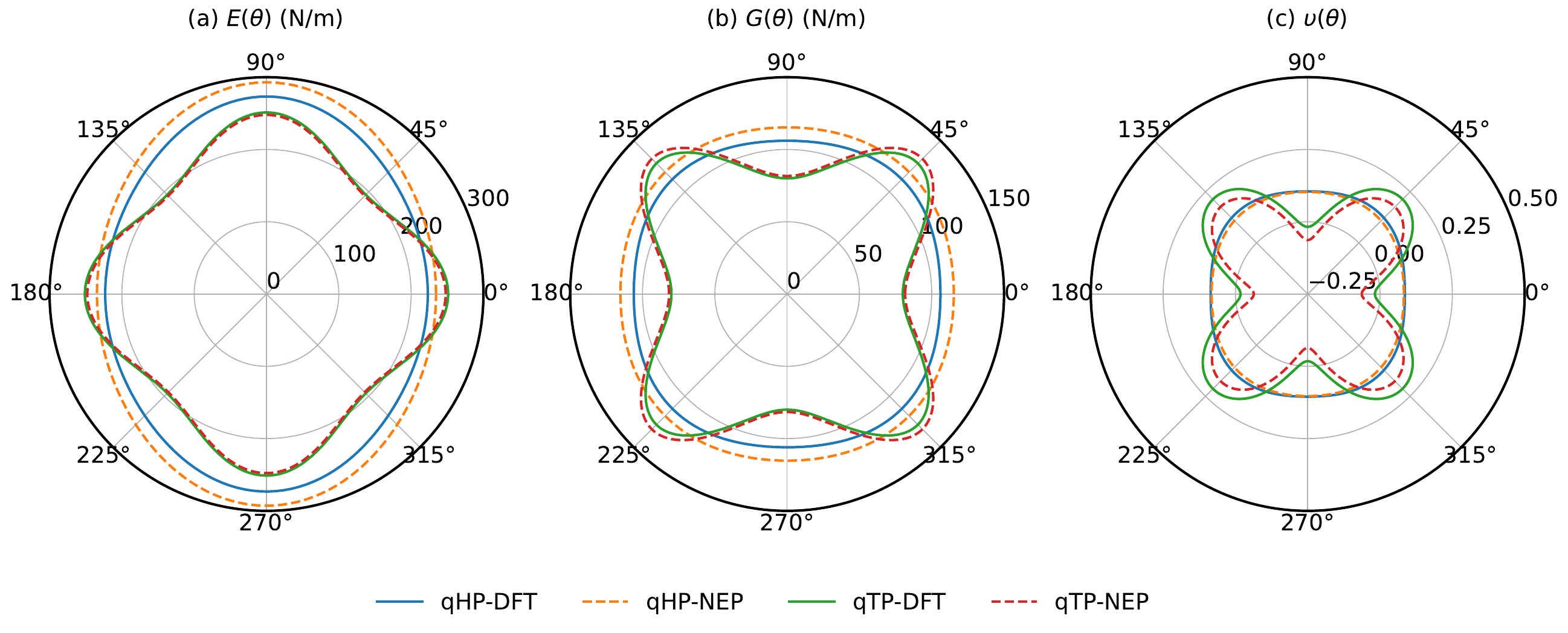} 
    \caption{Orientation-dependent (a) Young’s modulus \( E(\theta) \), (b) shear modulus \( G(\theta) \), and (c) Poisson’s ratio \( \nu(\theta) \) of monolayer \gls{qhp} and \gls{qtp} from \gls{dft} and \gls{nep} calculations, where the dashed lines represent \gls{nep} results and the solid lines represent \gls{dft} results.}
    \label{figelastic} 
\end{figure*} 

These calculated elastic constants further verify the mechanical stability of the \gls{qhp} and \gls{qtp} C$_{24}$ monolayers, as they satisfy the Born criteria for two-dimensional (2D) rectangular lattices~\cite{Born_1940,Mazdziarz20192dmater}, $C_{11} + C_{22} \pm \sqrt{4C_{12}^2 - (C_{11} - C_{22})^2} > 0$, $C_{66} > 0$. Furthermore, these elastic constants were used to estimate the orientation-dependent Young’s modulus $E (\theta)$, shear modulus $G (\theta)$ and Poisson’s ratio $\nu (\theta)$ according to the following equations~\cite{jaslukiewicz2008pssb, ying2022tension}:
\begin{align*}
E(\theta) &= \frac{1}{S_{11}a^4 + S_{22}b^4 + (S_{66} + 2S_{12})a^2b^2}, \\
G(\theta) &= \frac{1}{4\left[(S_{11} + S_{22} - 2S_{12})a^2b^2 + S_{66}(a^2 - b^2)^2\right]}, \\
\nu(\theta) &= -E(\theta)\left[(S_{11} + S_{22} - S_{66})a^2b^2 + S_{12}(a^4 + b^4)\right].
\end{align*}
where \( a = \cos(\theta) \) and \( b = \sin(\theta) \), with \( \theta \) being the orientation angle defined with respect to the \( x \)-axis. \( S_{11} \), \( S_{22} \), \( S_{12} \), and \( S_{66} \) are compliance constants, which can be obtained by inverting the elastic constant matrix as shown in \autoref{table:Cij}.

The obtained \( E(\theta) \), \( G(\theta) \), and in-plane Poisson's ratio \( \nu(\theta) \) are shown in \autoref{figelastic}. For the \gls{qhp} C$_{24}$ monolayer, all elastic properties exhibit pronounced anisotropy, particularly the $E(\theta)$. Specifically, $E(\theta)$ gradually increases as $\theta$ varies from \ang{0} to \ang{90},  with $E_x$ ($\theta =\ang{0}$, 223.1 N/m predicted by \gls{dft}) smaller than $E_y$ ($\theta =\ang{90}$, 273.2 N/m predicted by \gls{dft}). This trend is consistent with the relatively weaker diagonal interchain single bonds along the $x$ direction. 

In contrast, \gls{qtp} C$_{24}$ monolayer shows identical elastic properties along the $x$ and $y$ directions, due to its $S_4$ (fourfold improper rotation) symmetry. Interesting, along the $\ang{45}$, the Young’s modulus $E$ reaches its minimum, whereas the shear modulus $G$ attains its maximum. The in-plane Poisson’s ratio $\nu (\theta)$ changes from $-0.018$ along the $x$ and $y$ directions to a maximum positive value of 0.21 at $\theta =\ang{45}$, indicating an almost vanishing transverse strain under uniaxial loading along the principal axes. This behavior suggests that the monolayer \gls{qtp} C$_{24}$ is a promising zero Poisson's ratio  material for direction-specific mechanical sensors and actuators~\cite{huang2016negative}.

Comparatively, \gls{qhp} C$_{24}$ monolayer possesses a higher \( E_y \) (273.2 N/m predicted by \gls{dft}) and lower \( E_x \) (223.1 N/m predicted by \gls{dft}) compared with \gls{qtp} C$_{24}$ (251.2 N/m predicted by \gls{dft}) monolayer. This anisotropy of \gls{qhp} C$_{24}$ can be attributed to its more densely packed configuration and a higher concentration of interfullerene bonds along the $y$ direction. Interestingly, the Young’s moduli of both C$_{24}$ monolayers are approximately 1.5 times higher than those of their C$_{60}$ counterparts (see \autoref{table:Cij})~\cite{peng2023nanolett, ying2023extreme}. This enhancement arises from the triple noncoplanar bonds in the C$_{24}$ monolayers, in contrast to the [2 + 2] cycloaddition bonds in the polymerized C$_{60}$ sheets. The smaller molecular size of C$_{24}$ also leads to a higher density of interfullerene bonds, further increasing the overall stiffness. This is also consistent with the fact that the \gls{qtp} $C_{60}$ has been shown to be unstable in monolayer form based on both experimental~\cite{hou2022nature} and theoretical~\cite{ying2023extreme,peng2023nanolett,ying2025advances} studies, while \gls{qtp} $C_{24}$ monolayer has been predicted to be stable~\cite{wu2025jacs}, which is also confirmed by our direct \gls{md} simulations at finite temperatures (\autoref{figgp}) as well as the phonon dispersions (\autoref{figphonon}) and the Born stability criteria evaluated from the elastic constants (\autoref{table:Cij}).

Furthermore, the Poisson’s ratio of the \gls{qhp} C$_{24}$ monolayer is smaller than that of \gls{qhp} C$_{60}$ and several other 2D carbon allotropes, including graphene ($\nu \approx 0.30$)~\cite{jiang2016nanolett}, graphyne ($\nu \approx 0.42$)~\cite{kang2019acsami}, and the biphenylene network ($\nu \approx 0.31$)~\cite{luo2021scirep}. Specifically, the Poisson’s ratio of qHP C$_{60}$ typically falls within the range of 0.12--0.20~\cite{ying2023extreme}, whereas that of qHP C$_{24}$ remains in the lower range of 0.09--0.11. This relatively low $\nu$ reflects a stronger resistance to transverse deformation under uniaxial strain, further highlighting the mechanical robustness of \gls{qhp} C$_{24}$. 

Since phonon transport is strongly governed by lattice stiffness and bonding characteristics, such elastic anisotropy is expected to influence thermal transport as well. We therefore proceed to investigate the thermal conductivity and phonon-related thermal properties of the two monolayers.

\subsection{Thermal transport properties}
Using the \gls{nep}-C$_{24}$ potential, we performed \gls{hnemd} simulations to investigate the thermal transport properties of the \gls{qhp} and \gls{qtp} C$_{24}$ monolayers. For the \gls{qhp} C$_{24}$ monolayer, the thermal conductivities were calculated along both the $x$- and $y$- directions, using a rectangular simulation cell with a lateral dimensions of approximately $21.8\ \mathrm{nm} \times 21.6\ \mathrm{nm}$ that contains 31,920 atoms and an effective thickness of $0.629\ \mathrm{nm}$. For \gls{qtp} C$_{24}$ monolayer, we employed a square cell of approximately $24.4\ \mathrm{nm} \times 24.4\ \mathrm{nm}$ that contains 38,400 atoms with an effective thickness of $0.568\ \mathrm{nm}$. The effective thicknesses were chosen as the interlayer spacings of the AA-stacked bulk structures after structural optimization using \gls{dft}.

\begin{figure*}[htbp]
    \centering
    \includegraphics[width=2\columnwidth]{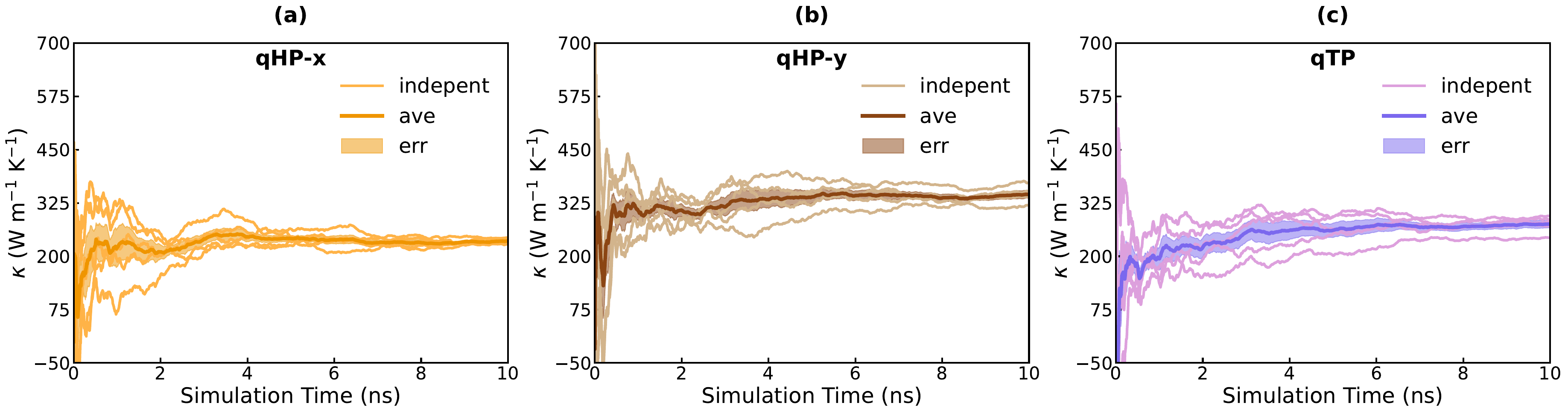}
    \caption{Running thermal conductivity as a function of simulation time for (a-b) the \gls{qhp} C$_{24}$ monolayer along the (a) $x$-direction and  (b) $y$-direction, and (c) \gls{qtp} monolayer at 300 K, calculated using the \gls{hnemd} method with the \gls{nep}-C$_{24}$ potential. In each panel, thin lines represent results from five individual runs (each with a 10 ns production time), while the thick solid line and the shaded region represent their average and the corresponding error bounds, respectively.}
    \label{figthermal}
\end{figure*}

\begin{table}[htb]
\centering
\setlength{\tabcolsep}{2.5mm}
\caption{Thermal conductivity $k$ (W m$^{-1}$ K$^{-1}$) of monolayer carbon allotropes.}
\label{table:k}
\begin{tabular}{llllll} %
\hline
Phase & Method & $k$ \\
\hline
\gls{qhp}-$x$ C$_{24}$ (This work)  & HNEMD  & 233(5)  \\
\gls{qhp}-$y$ C$_{24}$ (This work)  & HNEMD  & 341(9)  \\
\gls{qtp} C$_{24}$ (This work)   & HNEMD  & 272(9)   \\
\gls{qhp}-$x$ C$_{60}$~\cite{dong2023ijhmt}& HNEMD  & 102(3)  \\
\gls{qhp}-$y$ C$_{60}$~\cite{dong2023ijhmt}& HNEMD  & 107(7)  \\
BPF~\cite{dong2023ijhmt}& HNEMD  & 0.45(5)\\
graphene~\cite{Wu2024JChemPhys} &HNEMD &1855(56)\\
$\alpha$-graphyne~\cite{Yang2018pccp} & BTE & 21.11   \\
biphenylene-$x$~\cite{Veeravenkata2021Carbon} & BTE &  166\\
biphenylene-$y$~\cite{Veeravenkata2021Carbon} & BTE &  254\\

\hline
\end{tabular}
\end{table}

In all the \gls{md} simulations, the system was first equilibrated for $1\ \mathrm{ns}$ in the NPT (isothermal-isobaric) ensemble at 300 K using the Bussi-Donadio-Parrinello thermostat~\cite{bussi2007canonical} and a lateral pressure of 0 GPa controlled by a stochastic cell rescaling barostat~\cite{bernetti2020pressure}. This was followed by a $10\ \mathrm{ns}$ production run in the NVT (canonical) ensemble at 300 K with a Nos\'{e}--Hoover chain thermostat~\cite{tuckerman2010statistical}. A time step of 1 fs was used throughout, with periodic boundary conditions applied in the in-plane directions and a non-periodic boundary condition along the out-of-plane direction. Based on convergence tests, the driving force parameter was set to \( F_{\rm e} = 5 \times 10^{-6} \, \text{Å}^{-1} \) for all \gls{hnemd} simulations to ensure a stable nonequilibrium state and reliable heat flux. For each case, five independent simulations were performed. The thermal conductivity was computed as the ensemble average, with the corresponding standard error of the mean reported as statistical error.

As shown in \autoref{figthermal}, at $300\ \mathrm{K}$ the thermal conductivity of the \gls{qhp} C$_{24}$ monolayer along the $x$-direction was $233\pm 5$ W m$^{-1}$ K$^{-1}$, while that along the $y$-direction was $341\pm 9$ W m$^{-1}$ K$^{-1}$ confirming pronounced thermal anisotropy induced by the directional bonding configuration in the \gls{qhp} lattice. For the \gls{qtp} C$_{24}$ monolayer, the thermal conductivity is $272 \pm 9$ W m$^{-1}$ K$^{-1}$, falling between the $x$ and $y$ direction values of the \gls{qhp} monolayer.

\begin{figure}[htbp]
    \centering
    \includegraphics[width=0.48\textwidth]{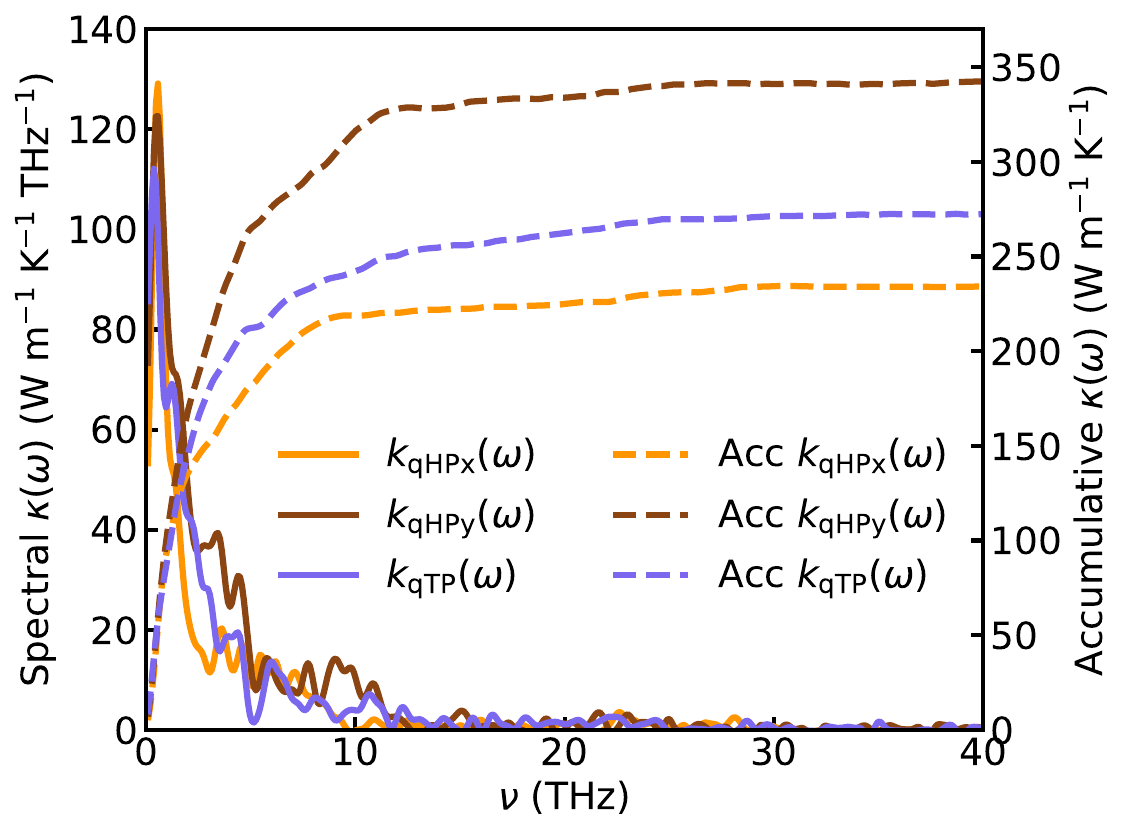} 
    \caption{Spectral thermal conductivity as a function of vibrational frequency for the \gls{qhp} C$_{24}$ monolayer (along the \(x\) and \(y\) directions) and the \gls{qtp} C$_{24}$ monolayer, calculated using the \gls{nep}-C$_{24}$ potential. Solid lines indicate spectral thermal conductivity (left axis), while dashed lines show the cumulative contribution (right axis).}
    \label{figshc} 
\end{figure}

\begin{figure*}[htbp]
    \centering
    \includegraphics[width=0.8\textwidth]{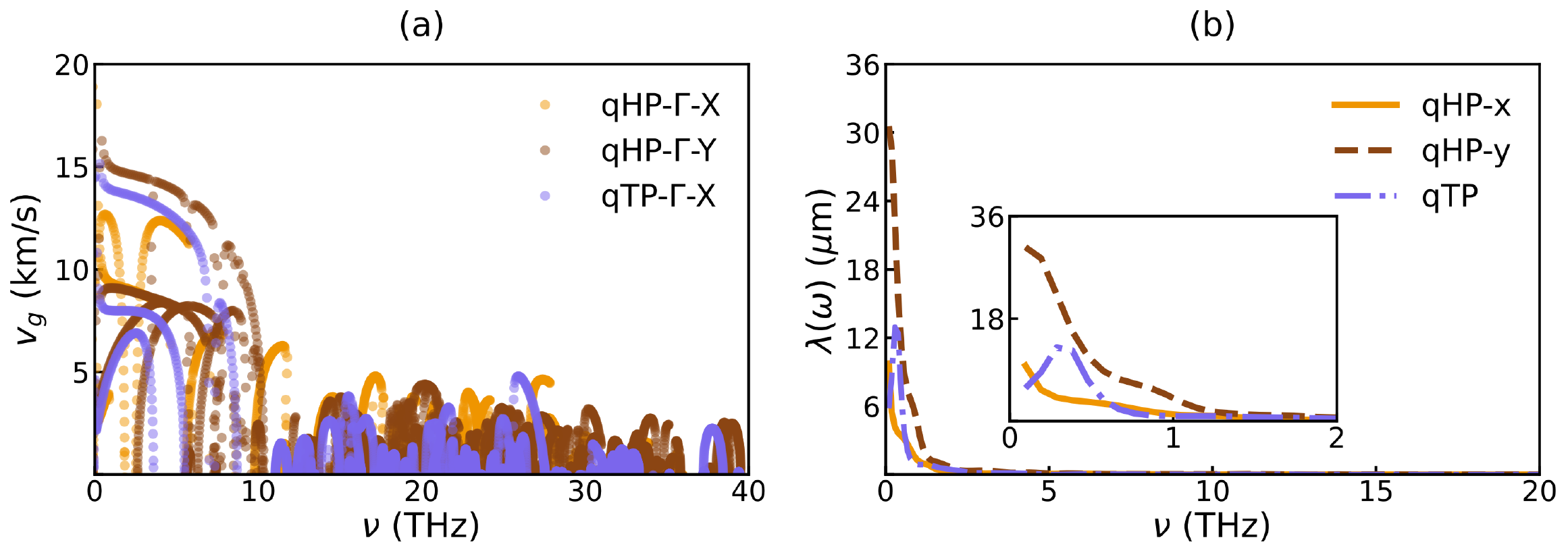} 
    \caption{(a) Phonon group velocity and (b) phonon mean free path \(\lambda(\omega)\)  as functions of frequency for \gls{qhp} monolayer in the \(x\) and \(y\) directions and for \gls{qtp} monolayer, calculated using the \gls{nep}-C$_{24}$ potential. The inset in (b) shows an enlarged view of the low-frequency region (\(\omega/2\pi < 2\) THz).}
    \label{figgp} 
\end{figure*} 

\begin{figure*}[htbp]
\begin{center}
\includegraphics[width=2\columnwidth]{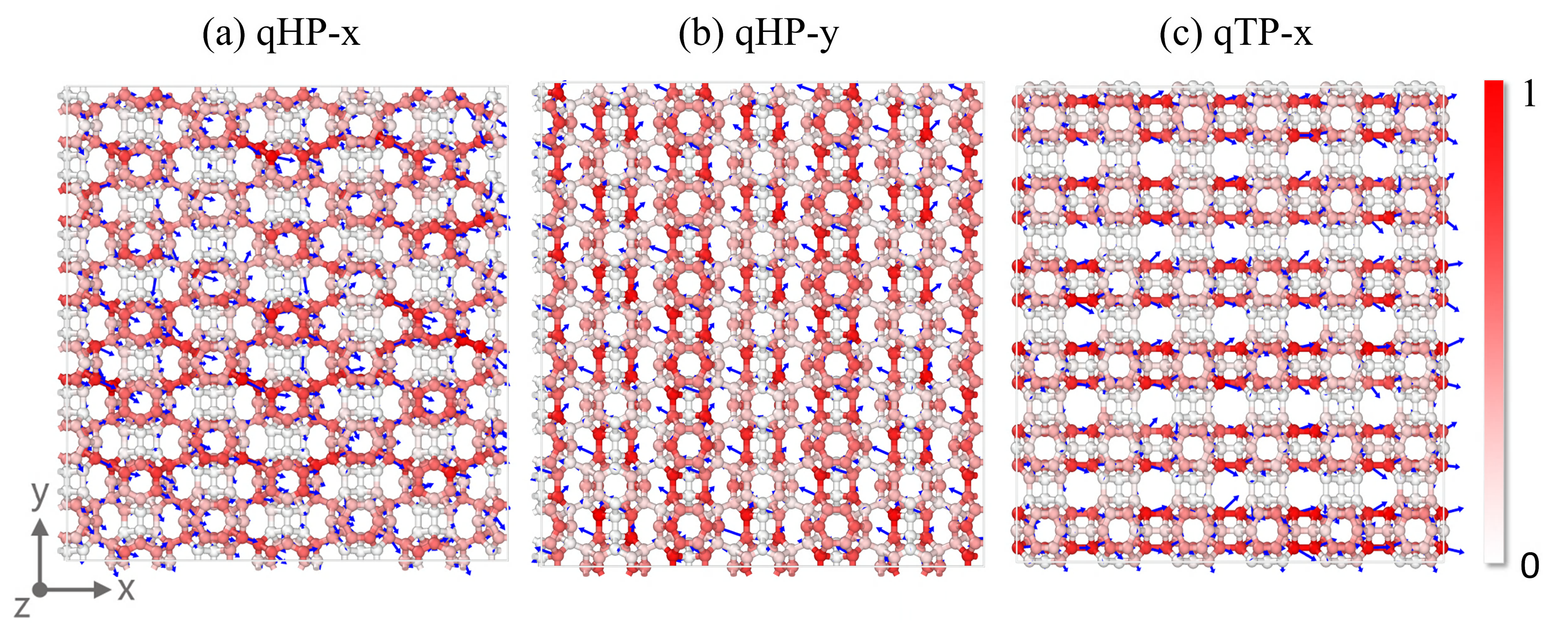}
\caption{Per-atom heat current distribution for (a) \gls{qhp} monolayer with heat transport in the $x$ direction, (b) \gls{qhp} monolayer with heat transport in the $y$ direction, and (c) \gls{qtp}, calculated using the \gls{nep}-C$_{24}$ potential. The color on the atoms indicates the normalized magnitude of the per-atom heat current in the transport direction, while the arrows represent both its magnitude and direction. For clarity, only a portion of the systems containing 31,920 atoms (\gls{qhp}) and 38,400 atoms (\gls{qtp}) is shown.}
\label{figjp}
\end{center}
\end{figure*}

To gain deeper insight into the microscopic phonon transport mechanisms, \autoref{figshc} shows the spectral thermal conductivity $\kappa(\omega)$ of \gls{qhp} (along both the $x$ and $y$ directions) and \gls{qtp} C$_{24}$ monolayers. For all the materials and transport directions considered, the dominant contribution to heat conduction originates from phonons with frequencies below approximately $\nu = \omega / 2\pi = 5~\mathrm{THz}$. This frequency range closely corresponds to that of the acoustic phonon branches in their respective phonon dispersion relations (see \autoref{figphonon}). These results clearly indicate that, in both \gls{qhp} and \gls{qtp} monolayers, heat transport is primarily governed by acoustic phonons. From the accumulative $\kappa(\omega)$, the thermal conductivity essentially saturates by about 10 THz. At saturation, the accumulative $\kappa(\omega)$ of \gls{qhp} monolayer along $y$ is the largest, that of \gls{qtp} is intermediate, and that of \gls{qhp} along $x$ is the smallest. This shows that the enhanced thermal conductivity of \gls{qhp} along $y$ mainly stems from stronger low-frequency (acoustic) phonon contributions across the low-frequency bands. 

This trend correlates with the phonon group velocities (see \autoref{figgp}(a)): in the low-phonon-frequency regime ($\nu < 10~\mathrm{THz}$), the group velocity of \gls{qhp} monolayer along $y$ (the $\Gamma$--$Y$ path) is significantly higher than that along $x$ (the $\Gamma$--$X$ path), while the \gls{qtp} monolayer values lie between these two. This behavior is consistent with the anisotropy in Young’s modulus of the \gls{qhp} monolayer (\autoref{figelastic}(a)): the stiffer $y$-direction, reinforced by parallel triple inter-fullerene bonds (larger $E_y$), supports faster acoustic phonons and thus higher thermal conductivity, whereas the softer $x$-direction with diagonal single bonds (smaller $E_x$) exhibits reduced group velocities and weaker heat transport. In contrast, \gls{qtp} C$_{24}$ shows identical elastic properties along the principal $x$ and $y$ axes, with Young’s modulus lying between the two directions of its \gls{qhp} counterpart, leading to a thermal conductivity that consistently falls between the $x$ and $y$ direction values of the \gls{qhp} monolayer. This picture is further supported by the calculated phonon \glspl{mfp} ($\lambda$) for \gls{qhp} and \gls{qtp} C$_{24}$ monolayers, which generally follow the hierarchy $\lambda_{\text{\gls{qhp}-y}} > \lambda_{\text{\gls{qtp}}} > \lambda_{\text{\gls{qhp}-x}}$ (see \autoref{figgp}(b)).

Beyond phonon group velocity and \gls{mfp}, additional microscopic insight can be obtained from the real-space heat current distribution, as illustrated in \autoref{figjp}. For the \gls{qhp} monolayer, when heat flow along the $x$ direction is mainly carried through the C-C single bonds that connect neighboring C$_{24}$ units, whereas along the $y$ direction it is dominanted by the three noncoplanar bonds aligned with the [010] direction. In the \gls{qtp} monolayer, due to its $S_4$ symmetry, inter-molecular heat transfer proceeds predominantly through the same three noncoplanar bonds for transport in both the $x$ and $y$ directions. Interestingly, in both phases, the central bond among the three inter-fullerene noncoplanar bonds appears to carry insignificant heat current.

Overall, these real space observations further demonstrate that, in both \gls{qhp} and \gls{qtp} C$_{24}$ monolayers, phonon transport is primarily mediated by strong covalent bonds rather than weak van der Waals interactions, consistent with our previous findings for the \gls{qhp} C$_{60}$ monolayer~\cite{dong2023ijhmt}. Within the consistent computational framework, we note that the thermal conductivities of \gls{qhp} and \gls{qtp} C$_{24}$ monolayers (234--341 W m$^{-1}$ K$^{-1}$) are roughly twice as those of \gls{qhp} C$_{60}$ monolayer (102--138 W m$^{-1}$ K$^{-1}$) and a few orders of magnitude higher than bulk-phase fullerene (0.45 W m$^{-1}$ K$^{-1}$). 
In a broader context, we note that the thermal conductivities of the fullerene monolayers are of the same order as those in biphenylene~\cite{ying2022thermal}, but are smaller than that of graphene (see~\autoref{table:k}).
We attribute this enhancement to the stronger inter-fullerene covalent bonds and the smaller molecular size of C$_{24}$, which together yield a higher density of inter-fullerene connections. These results highlight a ``small-is-stronger'' trend: the reduced fullerene size and distinct bonding topology play a decisive role in determining both the magnitude and anisotropy of thermal transport in fullerene network systems.

\section{Summary and Conclusions}

In summary, we have developed an accurate machine-learned potential, \gls{nep}-C$_{24}$, based on the neuroevolution potential framework to describe both \gls{qhp} and \gls{qtp} C$_{24}$ monolayers. The potential reproduces \gls{dft} results with excellent agreement in terms of energy, force, virial, radial distribution function, angular distribution function, and phonon dispersions, demonstrating its strong reliability for molecular dynamics simulations. Using this potential, we systematically investigated the mechanical and thermal transport properties of both phases. The \gls{qhp} monolayer exhibits pronounced in-plane anisotropy in its mechanical parameters and thermal conductivity, originating from its misaligned chain-like bonding topology and directional noncoplanar C–C covalent connections. In contrast, the \gls{qtp} monolayer displays nearly isotropic behavior owing to its symmetric bonding configuration. Homogeneous nonequilibrium molecular dynamics simulations combined with spectral decomposition further reveal that acoustic phonons below 5 THz dominate the heat conduction, and the anisotropic phonon group velocity and mean free path are responsible for the directional dependence of thermal transport in the qHP C$_{24}$ monolayer.

Our findings establish a direct correlation between bonding topology and heat transport anisotropy in fullerene-based \gls{2d} materials. In particular, a clear ``smaller is stronger'' effect emerges when comparing the elastic properties and thermal conductivities of C$_{24}$ network with their C$_{60}$ counterparts: the smaller C$_{24}$ fullerene network hosts a much higher density of inter-molecular connections due to its reduced molecule size and parallel triple inter-fullerene bonds, leading to enhanced stiffness and thermal transport. The \gls{nep} framework utilized herein is of general nature and can be readily extended to explore fullerene networks with defects, edge terminations, and varying nanoribbon widths. Our approach offers a powerful tool for the rational design of fullerene-based networks with tunable thermal conductivity and mechanical anisotropy, leveraging fullerene size and inter-fullerene bonding topology as design parameters.

\section*{Data availability}
All input and output files related to the training of the \gls{nep}-C$_{24}$ potential are freely available via the Zenodo link: https://doi.org/10.5281/zenodo.17562701.

\section*{Acknowledgments}
This work was supported by the National Science and Technology Advanced Materials Major Program of China (No. 2024ZD0606900). ZF is supported by the Science Foundation from Education Department of Liaoning Province
(No. LJ232510167001). We thank Ke Xu, Shujun Zhou and Zihan Tan for helpful discussions.

\end{document}